\documentclass[english]{extarticle}
\usepackage[T1]{fontenc}
\usepackage[latin9]{inputenc}
\usepackage{geometry}
\usepackage{array}
\usepackage{float}
\usepackage{units}
\usepackage{mathrsfs}
\usepackage{amsmath}
\usepackage{amsthm}
\usepackage{amssymb}
\usepackage{graphicx}

\makeatletter

\providecommand{\tabularnewline}{\\}

\numberwithin{equation}{section}
\numberwithin{figure}{section}

\DeclareMathSizes{12}{17.28}{9}{7}

\usepackage{braket}
\usepackage{remreset}

\@removefromreset{equation}{section}

\@removefromreset{figure}{chapter}

\AtBeginDocument{
  
}

\makeatother

\usepackage{babel}
\begin{document}

\title{In Wigner phase space, convolution explains why the vacuum majorizes
mixtures of Fock states}

\author{Luc Vanbever\date{%
    Ecole polytechnique de Bruxelles,\\%
    CP 165, Université libre de Bruxelles, 1050 Brussels, Belgium\\[3ex]%
    August 2021
}}
\maketitle
\begin{abstract}
I show that a nonnegative Wigner function that represents a mixture
of Fock states is majorized by the Wigner function of the vacuum state.
As a consequence, the integration of any concave function over the
Wigner phase space has a lower value for the vacuum state than for
a mixture of Fock states. The Shannon differential entropy is an example
of such concave function of significant physical importance. I demonstrate
that the very cause of the majorization lies in the fact that a Wigner
function is the result of a convolution. My proof is based on a new
majorization result dedicated to the convolution of the negative exponential
distribution with a precisely constrained function. I present a geometrical
interpretation of the new majorization property in a discrete setting
and extend this relation to a continuous setting. Findings presented
in this article might be expanded upon to explain why the Wigner function
of the vacuum majorizes - beyond mixtures of Fock states - many other
physical states represented by a nonnegative Wigner function.
\end{abstract}

\section{Introduction\label{sec:Introduction}}

Phase space representations were conceived almost ninety years ago
by Eugene Wigner \cite{Wigner-Review-2018}. Although the original
application was the description of states and systems in Quantum mechanics,
the Wigner function is today a successful tool in areas as diverse
as the study of radiation and optical systems~\cite{Optical_Systems},
quantum electronics, quantum chemistry or signal analysis \cite{Wigner-Review-2018}.
The Wigner phase space has also proved to be an efficient framework
in the analysis of both stationary and time-dependent systems.

The Wigner function keeps providing interesting insight when analysing
quantum systems. This is also strengthened by the fact that the Wigner
function has a series of - almost directly visible - features that
nicely reflect the physics of the quantum states they represent. For
instance, negative parts in the Wigner functions are viewed as the
signature of a non-classical behaviour \cite{Kenfack_2004}, whereas
in the absence of negative areas, it behaves as an actual probability
distribution\footnote{Note however that squeezed Gaussian states have a nonnegative Wigner
function but exhibit some non-classical behaviour.}. Another illustration is the normalization of the Wigner function
that simply mirrors the unit trace of the density operator.

The entropy of the Wigner function - and more specifically Shannon
differential entropy - is a quantity that relates to the amount of
information that can be gained on a quantum system. When entropy relations
involve the two variables of the two-dimensional Wigner phase space,
they can reflect the limited amount of knowledge that can be obtained
simultaneously on both variables. For instance, some lower bounds
on the sum of the entropies of the two phase space variables imply
Heisenberg's uncertainty relations \cite{Bia?ynicki-Birula1975,Hertz2}.

It has been conjectured that the joint entropy of the Wigner function
of a coherent state is the minimum joint entropy that any physical
state can have if it is represented by a nonnegative Wigner function~\cite{Hertz2}.
Visually, this would mean that a Wigner function can never be ``more
peaked'' than the Wigner function of a coherent state.

In this article, I prove this conjecture for the case of mixtures
of Fock states. I use the theory of majorization and I explain why
the Wigner function of the vacuum state majorizes the Wigner function
of mixtures of Fock states - in the cases where this Wigner function
is nonnegative. Such majorization relation directly implies entropy
(and seminorms) comparisons between the various distributions.

In section \ref{sec:Definitions and Properties}, I start by rewriting
a couple of definitions and properties that I use in this article.
The seven parts of my proof are presented in section \ref{sec:Proof}.
Section \ref{sec:Proofs Theorems 1=0000262} details the proof of
the two main theorems of this article and I conclude in section \ref{sec:Conclusion}.

\section{Definitions and Properties\label{sec:Definitions and Properties}}

In this article, a vector $x$ is said to be nonnegative if all its
elements are nonnegative $x_{i}\geqslant0,\:\forall i$. A vector
is said to be normalized if the sum of its elements is equal to $1$.
Similarly, in the continuous case, a function $f\left(x\right)$ defined
on the measure space $\left(H,\mathscr{I},\nu\right)$ is said to
be nonnegative if it takes nonnegative values on its domain $H$.
A function $f\left(x\right)$ is said to be normalized if
\[
\intop f(x)\;\mathrm{d}\nu(x)=1,
\]
where the integration is performed over the complete domain $H$.

\subsection*{Some Definitions of Majorization\label{sec:Definitions of Majorization}}

I first recall a couple of equivalent definitions for majorization,
both in a discrete setting and in a continuous setting. I restrict
the description of the equivalent definitions to the ones that will
be used in this article.

In a discrete setting, if $x$,$\:y\in\mathbb{R}^{N}$, the following
five statements are equivalent \cite{Hickey_1983,Marshall&Olkin}:
\begin{enumerate}
\item The vector $x$ majorizes $y$, denoted $x\succ y$.
\item \begin{subequations}
\begin{eqnarray}
\sum_{i=1}^{k}x_{i}^{\downarrow}\geqslant\sum_{i=1}^{k}y_{i}^{\downarrow}, &  & \qquad k=1,\dots,N-1,\qquad\text{and}\label{eq:Definition Discrete-ReOrdered Sum}\\
\sum_{i=1}^{N}x_{i}=\sum_{i=1}^{N}y_{i},\label{eq:Definition Discrete Total Sum}
\end{eqnarray}
\end{subequations}where the vectors $x{}^{\downarrow}$ and $y{}^{\downarrow}$
include the same elements, respectively as $x$ and $y$, but re-sorted
in a non-increasing order.
\item 
\begin{eqnarray}
\sum_{i=1}^{N}\Phi\left(x_{i}\right)\geqslant\sum_{i=1}^{N}\Phi\left(y_{i}\right),\label{eq:Definition Discrete-Convex Functions}
\end{eqnarray}
 for all continuous convex functions $\Phi:\mathbb{R}\longrightarrow\mathbb{R}$.
\item There exists a $N\times N$ doubly stochastic matrix $D$ such that
\begin{eqnarray}
y=Dx.\label{eq:Definition Discrete-Doubly Stochastic}
\end{eqnarray}

\item The vector $y$ can be derived from $x$ by performing a limited number
of Robin Hood transfers. A transfer of a positive amount $\Delta$
from $x{}_{i}$ to $x{}_{j}$, giving $y_{i}=x_{i}-\Delta$ and $y_{j}=x_{j}+\Delta$
is said to be a Robin Hood transfer if $\Delta\leqslant x_{i}-x_{j}$.
This definition is valid for nonnegative vectors $x$ and $y$. A
Robin Hood transfer is also called a Dalton transfer or sometimes
a ``pinch'' \cite{Marshall&Olkin}.
\end{enumerate}
Similarly, in a continuous setting, if $f$ and $g$ are nonnegative
$\nu$-integrable functions defined on the measure space $\left(H,\mathscr{I},\nu\right)$,
the following four statements are equivalent \cite{hickey_1984,Marshall&Olkin}:
\begin{enumerate}
\item The function $f$ majorizes $g$, denoted $f\succ g$.
\item \begin{subequations}
\begin{eqnarray}
\intop_{0}^{t}f_{\downarrow}(u)\;\mathrm{d}\nu(u)\geqslant\intop_{0}^{t}g_{\downarrow}(u)\;\mathrm{d}\nu(u), &  & \qquad\forall t\in\left[0,\nu\left(H\right)\right)\quad\text{and}\label{eq:Definition Continuous-ReArranged Sum}\\
\intop f(u)\;\mathrm{d}\nu(u)=\intop g(u)\;\mathrm{d}\nu(u),\label{eq:Definition Total-Sum}
\end{eqnarray}
\end{subequations}where $f_{\downarrow}$ (and identically for $g_{\downarrow}$)
denotes the decreasing rearrangement of $f$ defined in two steps:\begin{subequations}
\begin{eqnarray}
m_{f}(t)=\nu\left(\left\{ x:f\left(x\right)>t\right\} \right), &  & \qquad t\geqslant0,\qquad\text{and},\label{eq:Measure1}\\
f_{\downarrow}(u)=\sup\left\{ t:m_{f}(t)>u\right\} , &  & \qquad0\leqslant u\leqslant\nu\left(H\right).\label{eq:Measure2}
\end{eqnarray}
\end{subequations}
\item 
\begin{eqnarray}
\intop\Phi(f)\;\mathrm{d}\nu\geqslant\intop\Phi(g)\;\mathrm{d}\nu,\label{eq:Definition Continuous-Convex Functions}
\end{eqnarray}
for all continuous convex functions $\Phi:\mathbb{R}\longrightarrow\mathbb{R}$
with $\Phi\left(0\right)=0$, for which the integrals exist.
\item 
\begin{equation}
g(y)=\intop k\left(x,y\right)\:f\left(x\right)\:\mathrm{d}\nu\left(x\right)\label{eq:Definition Continuous Doubly Stochastic}
\end{equation}
where $k:\:H\times H\longrightarrow\left[0,+\infty\right)$ is a doubly
stochastic function, that is, $\intop k\left(x,y\right)\:\mathrm{d}\nu\left(x\right)=1$
for all $y\in H$ and $\intop k\left(x,y\right)\:\mathrm{d}\nu\left(y\right)=1$
for all $x\in H$. Note that some authors present this condition as
sufficient and not as an equivalent definition \cite{hickey_1984}.
In this article, Eq.~(\ref{eq:Definition Continuous Doubly Stochastic})
is only used as a sufficient condition for majorization.
\end{enumerate}
A number of properties of discrete majorization extend to continuous
majorization and we note the similarities between the equivalent definitions
in the discrete and in the continuous settings (not all are reproduced
here). In the continuous setting, some of the equivalent definitions
are applicable to nonnegative functions; this nonnegativity restriction
can be relaxed if $\nu\left(H\right)$ is finite \cite{Marshall&Olkin}.
I keep this restriction as the support of the functions studied in
this article is not finite.

\subsection*{Majorization Property of the Convolution of Two Probability Distributions\label{sub:Majorization Property of...}}

The convolution vector of two probability distributions is majorized
by the two original probability distributions \cite{Hickey_1983}.
I particularize here the proof \cite{Hickey_1983,Marshall&Olkin}
to the case where the first probability vector has one more element
than the second probability vector. The discrete convolution of a
normalized vector $y=\left(y{}_{1},y{}_{2},\dots,y_{N+1}\right)^{\intercal}$
and a vector $x=\left(x{}_{1},x{}_{2},\dots,x_{N}\right)^{\intercal}$
is formulated as the multiplication by a Toeplitz matrix constructed
as follows:

\begin{eqnarray}
z=y\ast x=\left(\begin{array}{cccc}
y{}_{1} & 0 & \cdots & 0\\
y{}_{2} & y{}_{1} & \ddots & \vdots\\
\vdots & y{}_{2} & \ddots & 0\\
y{}_{N} & \vdots & \ddots & y{}_{1}\\
y{}_{N+1} & y{}_{N} & \ddots & y{}_{2}\\
0 & y{}_{N+1} & \ddots & \vdots\\
\vdots & \ddots & \ddots & y{}_{N}\\
0 & \cdots & 0 & y{}_{N+1}
\end{array}\right)\left(\begin{array}{c}
x{}_{1}\\
x{}_{2}\\
\vdots\\
x{}_{N}
\end{array}\right),\label{eq:Toeplitz convolution}
\end{eqnarray}

where the symbol $\ast$ denotes the convolution. The resulting probability
vector $z$ has a maximum of $2N$ non-zero elements. If we build
a $2N\times2N$ circulant matrix $C$ by adding $N$ columns at the
right of the above Toeplitz matrix, the probability vector $z$ can
also be written as

\begin{eqnarray}
z=C\left(\begin{array}{c}
x{}_{1}\\
x{}_{2}\\
\vdots\\
x{}_{N}\\
0\\
0\\
\vdots\\
0
\end{array}\right),\label{eq:Circulant convolution}
\end{eqnarray}
where $N$ zero elements are added to the vector $x$. With Eq.~(\ref{eq:Definition Discrete-Doubly Stochastic}),
the equality in Eq.~(\ref{eq:Circulant convolution}) proves that
$x\succ z$ because the circulant matrix $C$ is doubly stochastic.

These results have been generalized to the continuous distributions.
In particular, if $X$ and $Y$ are independent random variables with
absolutely continuous distribution functions, then the distribution
of $X+Y$ is majorized by both the distributions of $X$ and $Y$
\cite{hickey_1984} - the explicit proof is not provided in the referenced
article.

If $c,\:f:\:\mathbb{R}^{+}\longrightarrow\mathbb{R}$ are two functions
whose support is the positive real axis, the convolution of $c\left(x\right)$
and $f\left(x\right)$ is 
\begin{eqnarray}
g\left(x\right)=c\ast f=\intop_{0}^{x}c\left(t\right)f\left(x-t\right)\mathrm{d}t, &  & \qquad x\geqslant0.\label{eq:Continuous convolution}
\end{eqnarray}

A discrete version of this integral matches the formulation involving
the Toeplitz matrix of Eq.~(\ref{eq:Toeplitz convolution}).

\subsection*{The Wigner Function of the Fock States\label{sub:The Wigner Function of...}}

The definition of the Wigner function of the quantum state with density
operator $\hat{\rho}$ is \cite{leonhardt_2010}

\begin{eqnarray}
W(x,p)=\frac{1}{2\pi}\intop_{-\infty}^{+\infty}\:\langle x-\frac{x^{\prime}}{2}|\hat{\rho}|p+\frac{x^{\prime}}{2}\rangle\:e^{ix^{\prime}p}\:\mathrm{d}x^{\prime},\label{eq:Wigner Definition}
\end{eqnarray}
where $x$ and $p$ are canonically conjugate variables.

With this dimensionless definition, the Wigner function of the eigenstate
of the Number operator with integer eigenvalue $n$ reads \cite{leonhardt_2010}

\begin{eqnarray}
W_{n}(r,\theta)=\tilde{W}_{n}(r^{2})=\frac{\left(-1\right)^{n}}{\pi}\:L_{n}\left(2r^{2}\right)\:e^{-r^{2}},\label{eq:Fock State Definition}
\end{eqnarray}

where $r$ and $\theta$ are the polar coordinates in Wigner phase
space and $L_{n}\left(x\right)$ denotes the Laguerre polynomial of
the $n$-th order. The eigenstates of the Number operator are referred
to as Fock states and the state corresponding to $n=0$ as the vacuum
state.

\subsection*{A Set of States with Nonnegative Wigner Function \label{sub:A Set of States with Nonnegative Wigner Function}}

Some recent results about nonnegative Wigner functions will be used
in my proof. When injecting the Fock states $|m\rangle$ and $|n\rangle$
into the two inputs of a balanced beamsplitter, the states at each
of the two outputs of the balanced beamsplitter can be written as
the following mixture of Fock states \cite{QuantumWEntropy,Nielsen&Chuang}:

\begin{eqnarray}
\hat{\sigma}\left(m,n\right)=\left(m!\,n!\,2^{m+n}\right)^{-1}\:\sum_{z=0}^{m+n}\sum_{i=max\left(0,z-n\right)}^{min\left(z,m\right)}\;\sum_{j=max\left(0,z-n\right)}^{min\left(z,m\right)}\;\left(-1\right)^{i+j}\qquad\qquad\nonumber \\
\times\left(\begin{array}{c}
m\\
i
\end{array}\right)\left(\begin{array}{c}
n\\
z-i
\end{array}\right)\left(\begin{array}{c}
m\\
j
\end{array}\right)\left(\begin{array}{c}
n\\
z-j
\end{array}\right)\:z!\,\left(m+n-z\right)!\quad|z\rangle\langle z|.\label{eq:SigmaStates}
\end{eqnarray}

It was proved that the states $\hat{\sigma}\left(m,n\right)$ have
a nonnegative Wigner function \cite{QuantumWEntropy}. For more details
about the generation and properties of the states $\hat{\sigma}\left(m,n\right)$,
I refer the reader to the article \cite{QuantumWEntropy} and to the
articles cited therein.

In the present article, I will use Eq.~(\ref{eq:SigmaStates}) as
the definition of a parameterized set of states that have a nonnegative
Wigner function.

\section{Proof\label{sec:Proof}}

In this section, I prove that the Wigner function of the vacuum majorizes
the Wigner function of any mixture of Fock states when the latter
Wigner function is nonnegative. The proof consists of seven parts.

\subsection*{Definition 1 \label{sec:Definition 1}}

In the Wigner phase space, the Wigner function $W_{\alpha}$ majorizes
$W_{\beta}$, denoted as $W_{\alpha}\underset{W}{\succ}W_{\beta}$,
if 
\begin{eqnarray}
\intop\Phi(W_{\alpha}\left(\vec{s}\right))\;\mathrm{d}\vec{s}\;\geqslant\intop\Phi(W_{\beta}\left(\vec{s}\right))\;\mathrm{d}\vec{s},\label{eq:W-majorization definition}
\end{eqnarray}
for all continuous convex functions\footnote{A convex function of utmost interest is the opposite of the Shannon
differential entropy, for which $\Phi\left(W\left(\vec{s}\right)\right)=W\left(\vec{s}\right)\log\left(W\left(\vec{s}\right)\right)$
with the Wigner function $W\left(\vec{s}\right)$ assumed nonnegative.} $\Phi:\mathbb{R}\longrightarrow\mathbb{R}$ with $\Phi\left(0\right)=0$,
for which the integrals over the complete $2$-dimensional phase space
exist.

For Wigner functions that do not depend on the angular coordinate
but only depend on the radial coordinate $r$, the majorization in
Eq.~(\ref{eq:W-majorization definition}) is an application of the
definition in Eq.~(\ref{eq:Definition Continuous-Convex Functions})
with the measure $\nu=r^{2}$. I use the index $W$ below the majorization
symbol in $W_{\alpha}\underset{W}{\succ}W_{\beta}$ to refer to the
majorization with respect to the measure $\nu=r^{2}$.

In this article, I study convex combinations of Fock states and I
only consider Wigner functions that depend on $r^{2}$, as defined
in Eq.~(\ref{eq:Fock State Definition}). A direct application of
the definition in Eq.~(\ref{eq:W-majorization definition}) to two
such functions, $W_{\alpha}=W_{\alpha}\left(r^{2}\right)$ and $W_{\beta}=W_{\beta}\left(r^{2}\right)$
shows that
\begin{eqnarray}
W_{\alpha}\left(r^{2}\right)\underset{W}{\succ}W_{\beta}\left(r^{2}\right) & \iff & W_{\alpha}\left(z\right)\succ W_{\beta}\left(z\right),\label{eq:Link between W-majorization and majorization}
\end{eqnarray}
where the change of variable $z=r^{2}$ was performed. The majorization
on the right-hand side of Eq.~(\ref{eq:Link between W-majorization and majorization})
is the majorization with respect to the usual measure, that is, $\nu\left(z\right)=z$.
The domain of definition of the functions $W_{\alpha}\left(z\right)$
and $W_{\beta}\left(z\right)$ is $\left[0,+\infty\right)$. As an
example, to compare the Wigner function of the vacuum and the Wigner
function of an equally weighted mixture of the first two Fock states,
we can write

\begin{eqnarray}
W_{0}\underset{W}{\succ}\frac{1}{2}\left(W_{0}+W_{1}\right) & \iff & e^{-z}\succ ze^{-z},\label{eq:W-majorization example}
\end{eqnarray}
where $z\in\mathbb{R}^{+}$.\medskip{}

\subsection*{Formulation as a Convolution\label{sec:Formulation as Convolution}}

Thanks to Eq.~(\ref{eq:Link between W-majorization and majorization}),
to obtain a majorization relation between the vacuum and the Wigner
function $W_{\beta}\left(r^{2}\right)$ of a convex combination of
Fock states, we can compare $W_{\alpha}\left(z\right)=e^{-z}$ with
$W_{\beta}\left(z\right)$. To perform this comparison, it will prove
useful to formulate $W_{\beta}\left(z\right)$ as a convolution.

Eq.~(\ref{eq:Fock State Definition}) of a Fock state shows that
$W_{\beta}\left(z\right)$ is the product of a polynomial with $e^{-z}$
denoted here by $W_{\beta}\left(z\right)=P\left(z\right)e^{-z}$.

The expression $P\left(z\right)e^{-z}$ can also be viewed as a linear
combination of Erlang distributions of rate parameter equal to $1$.
The support of the Erlang distribution is $\mathbb{R}^{+}$ and its
general expression is

\begin{eqnarray}
E_{k+1}(z)=\frac{z^{k}e^{-z}}{k!}=p_{k}(z)e^{-z}, &  & \qquad k=0,1,2\dots\quad\text{and}\quad z\geqslant0,\label{eq:Erlang Definition}
\end{eqnarray}
which defines $p_{k}(z)$. The Erlang distribution of shape parameter
$k+1$ is the convolution as defined by Eq.~(\ref{eq:Continuous convolution}),
of an Erlang distribution of shape parameter $k$ with the negative
exponential distribution. If I also introduce the derivative of the
polynomial $p_{k}(z)$,  for $k\geqslant1$,

\begin{eqnarray}
p_{k}(z)e^{-z}=E_{k+1}(z)=E_{k}(z)\ast e^{-z}=\left(p_{k}^{\prime}(z)e^{-z}\right)\ast e^{-z}, &  & \quad z\geqslant0.\label{eq:Equality for polynomials}
\end{eqnarray}
$P\left(z\right)$ is a linear combination of the $p_{k}(z)$ and
Eq.~(\ref{eq:Equality for polynomials}) gives

\begin{eqnarray}
P\left(z\right)e^{-z}=\sum_{k=0}^{+\infty}a_{k}p_{k}(z)e^{-z}=a_{0}e^{-z}+\left(\sum_{k=1}^{+\infty}a_{k}p_{k}^{\prime}(z)e^{-z}\right)\ast e^{-z}, &  & \quad z\geqslant0.\label{eq:Equality2 for polynomials}
\end{eqnarray}
The function $W_{\beta}\left(z\right)$ can finally be formulated
as the following convolution:

\begin{eqnarray}
W_{\beta}\left(z\right)=P\left(z\right)e^{-z}=\left(a_{0}\delta\left(z\right)+P^{\prime}\left(z\right)e^{-z}\right)\ast e^{-z}, &  & \quad z\geqslant0,\label{eq:Wigner =00003D Convolution}
\end{eqnarray}
where the Dirac delta function takes the term $k=0$ of the sum into
account and $a_{0}$ was implicitly defined in Eq.~(\ref{eq:Equality2 for polynomials})
as the zero order coefficient of $P\left(z\right)$.

Note that the fact that $E_{k+1}(z)$ is obtained from the convolution
of the distributions $E_{k}(z)$ and $e^{-z}$ implies that $P\left(z\right)e^{-z}\prec e^{-z}$
in the cases where the coefficients of the polynomial $P\left(z\right)$
are nonnegative~\cite{hickey_1984}. This proves for instance that
$W_{0}\underset{W}{\succ}\frac{1}{2}\left(W_{0}+W_{1}\right)$.

Eq.~(\ref{eq:Wigner =00003D Convolution}) actually provides an indirect
representation of $W_{\beta}\left(r^{2}\right)$ as the convolution
of a continuous function (except in $z=0$) with the negative exponential
$e^{-z}$ that represents the vacuum.

\subsection*{Theorem 1 \label{sec:Theorem 1}}

Theorem 1 is a new majorization property applicable to the convolution
of a discrete negative exponential $v{}_{0}$ with a vector $x$ that
can have negative elements.

Let the non-normalized vector $v{}_{0}=\left(a^{0},a^{1},a^{2},\dots,a^{N-1}\right){}^{\intercal}$
with $0<a<1$ and let $x$ be a normalized $N+1$-dimensional vector
that can be written as the sum

\begin{equation}
x^{\intercal}=\left(1,0,\dots,0\right)+\sum_{k=1}^{N}\left(0,\dots,0,-\lambda_{k},\lambda_{k},0,\dots,0\right),\label{eq:G as convolution}
\end{equation}
where $\lambda_{k}\geqslant0,\;k=1,\dots,N$, and the $k$-th and
$k+1$-th elements are the only non-zero elements of each vector in
the sum of Eq.~(\ref{eq:G as convolution}).

Let $G$ be the $2N$-dimensional vector defined by the convolution
$G=x\ast v_{0}$. My main theorem states that
\begin{eqnarray}
\text{If }\;G\geqslant0\;\text{ Then }\;v_{0}\succ G.\label{eq:Main Theorem-Discrete}
\end{eqnarray}

The proof of Theorem 1 is provided in section \ref{sec:Proofs Theorems 1=0000262}.

\subsection*{Lemma 1 \label{sec:Lemma 1}}

A vector $x\in\mathbb{R}^{N+1}$ can be decomposed into a sum of the
form

\begin{equation}
x^{\intercal}=\left(\lambda_{0},0,\dots,0\right)+\sum_{k=1}^{N}\left(0,\dots,0,-\lambda_{k},\lambda_{k},0,\dots,0\right),\label{eq:Sum of minus/plus}
\end{equation}
where all $\lambda_{k}\geqslant0$, if and only if

\begin{equation}
\sum_{i=k}^{N}x_{i}\geqslant0,\qquad k=0,\dots,N.\label{eq:Sum of Last N}
\end{equation}

To first prove the inverse implication, it suffices to decompose the
vector $x$ into a sum of terms of the form $\left(0,\dots,0,-\lambda_{k},\lambda_{k},0,\dots,0\right)^{\intercal}$,
which also proves the unicity of the decomposition. As per Eq.~(\ref{eq:Sum of Last N}),
$x{}_{N}$ is positive and the last term of the sum must be $\left(0,\dots,0,-x_{N},x_{N}\right)^{\intercal}$.
Eq.~(\ref{eq:Sum of Last N}) directly shows that the $k$-th term
$\left(0,\dots,0,-\sum_{i=k}^{N}x_{i},\sum_{i=k}^{N}x_{i},0,\dots,0\right)^{\intercal}$
of the decomposition has the form required in Eq.~(\ref{eq:Sum of minus/plus}).
Finally, because $\lambda_{0}=\sum_{i=0}^{N}x_{i}$, the parameter
$\lambda_{0}\geqslant0$ and $\lambda_{0}=1$ if the vector $x$ is
normalized.

The direct implication derives from the fact that a single vector
of the form $\left(0,\dots,0,-\lambda_{k},\lambda_{k},0,\dots,0\right)^{\intercal}$
with $\lambda_{k}\geqslant0$ satisfies Eq.~(\ref{eq:Sum of Last N})
and that the sum of vectors that satisfies Eq.~(\ref{eq:Sum of Last N})
also satisfies that same equation.

\subsection*{Theorem 2 \label{sec:Theorem 2}}

Theorem 2 is the extension of Theorem 1 - combined with Lemma 1 -
to the continuous case.

Let $c(z)$ be a function defined on $\left[0,+\infty\right)$, integrable
and continuous on its domain, except in $z=0$ where one term of the
function $c(z)$ can be proportional to a Dirac delta function. The
function $c(z)$ also fulfills the conditions\vspace{-10bp}

\begin{subequations}
\begin{eqnarray}
\intop_{x}^{+\infty}\;c\left(z\right)\;\mathrm{d}z\geqslant0, & \qquad\forall\:x>0 & \qquad\text{and}\label{eq:Entry Partial Sum}\\
\intop_{0}^{+\infty}\;c\left(z\right)\;\mathrm{d}z=1.\label{eq:Entry Sum is One}
\end{eqnarray}
\end{subequations}Let $e^{-z}$ denote the negative exponential distribution
defined on $\left[0,+\infty\right)$ and let the function $g$ be
the convolution $g\left(z\right)=c\left(z\right)\ast e^{-z}$ as defined
by Eq.~(\ref{eq:Continuous convolution}). Theorem 2 states that

\begin{eqnarray}
\text{If }\;g\left(z\right)\geqslant0\;\text{ Then }\;e{}^{-z}\succ g\left(z\right).\label{eq:Main Theorem-Continuous}
\end{eqnarray}

The proof of Theorem 2 is provided in section \ref{sec:Proofs Theorems 1=0000262}.

\subsection*{Lemma 2 \label{sec:Lemma 2}}

The Wigner function of the equally weighted mixture of the first $M+1$
Fock states is nonnegative, for all $M\in\mathbb{N}$.

To prove this result, I consider the $M+1$ states $\hat{\sigma}\left(m,M-m\right),$
$m=0,\dots,M$. The definition of $\hat{\sigma}\left(m,n\right)$
was reproduced in Eq.~(\ref{eq:SigmaStates}) and I keep a constant
value for $M=m+n$,

\begin{eqnarray}
\hat{\sigma}\left(m,M-m\right)=\sum_{z=0}^{M}a_{m,z}|z\rangle\langle z|,\qquad m=0,\dots,M,\label{eq:MSigma States}
\end{eqnarray}
where

\begin{eqnarray}
a_{m,z}=2^{-M}\sum_{i=max\left(0,z+m-M\right)}^{min\left(z,m\right)}\sum_{j=max\left(0,z+m-M\right)}^{min\left(z,m\right)}\;\left(-1\right)^{i+j}\left(\begin{array}{c}
m\\
i
\end{array}\right)\left(\begin{array}{c}
M-m\\
z-i
\end{array}\right)z!\,(M-z)!\nonumber \\
\times\left(\begin{array}{c}
m\\
j
\end{array}\right)\left(\begin{array}{c}
M-m\\
z-j
\end{array}\right)/\left(m!\,(M-m)!\right).\label{eq:Coefficients Sigma}
\end{eqnarray}

We can notice that each term of the double sum in Eq.~(\ref{eq:Coefficients Sigma})
has a symmetry relatively to the variables $m$ and $z$:

\begin{equation}
\left(\begin{array}{c}
m\\
i
\end{array}\right)\left(\begin{array}{c}
M-m\\
z-i
\end{array}\right)z!\,(M-z)!=\left(\begin{array}{c}
z\\
i
\end{array}\right)\left(\begin{array}{c}
M-z\\
m-i
\end{array}\right)m!\,(M-m)!\:,\label{eq:Symmetry m=000026z}
\end{equation}

\medskip{}
and that the indices $i$ and $j$ also run through the same values
when $m$ and $z$ are swapped. This indicates that $a_{m,z}=a_{z,m}$
and the normalization $\sum_{z=0}^{M}a_{m,z}=1$ for $m=0,\dots,M$,
implies that summing on the index $m$ also gives the value $\sum_{m=0}^{M}a_{m,z}=1$
for $z=0,\dots,M$.

Hence, the mixture with equal weights of the states $\hat{\sigma}\left(m,M-m\right)$
provides\footnote{The physical reason for this result is not commented upon in this
article.}

\begin{eqnarray}
\frac{1}{M+1}\:\sum_{m=0}^{M}\hat{\sigma}\left(m,M-m\right)=\frac{1}{M+1}\:\sum_{z=0}^{M}|z\rangle\langle z|\,,\label{eq:MixtureOfSigmas}
\end{eqnarray}
which has a nonnegative Wigner function. This completes the proof
of Lemma 2.

\subsection*{Final Result \label{sec:Final Result}}

I have established in Eq.~(\ref{eq:Wigner =00003D Convolution})
that a function of the form $W_{\beta}\left(z\right)=P\left(z\right)e^{-z}$
can be expressed as the convolution of $a_{0}\delta\left(z\right)+P^{\prime}\left(z\right)e^{-z}$
with the negative exponential $e^{-z}$ distribution, where $a_{0}$
is the zero order coefficient of the polynomial $P\left(z\right)$.
The variable $z\in\:\mathbb{R}^{+}$ and the convolution is defined
by Eq.~(\ref{eq:Continuous convolution}).

To analyse mixtures of Fock states, I choose $P\left(z\right)=\left(-1\right)^{n}\;L_{n}\left(2z\right)$,
where $L_{n}\left(x\right)$ is the Laguerre polynomial of the $n$-th
order. The first entry condition of Theorem 2, Eq.~(\ref{eq:Entry Partial Sum}),
becomes

\begin{eqnarray}
\intop_{x}^{+\infty}\;\left(-1\right)^{n}\;L_{n}^{\prime}\left(2z\right)e^{-z}\mathrm{d}z\geqslant0,\qquad\forall x>0,\label{eq:Entry Condition with Laguerre}
\end{eqnarray}
where $L_{n}^{\prime}\left(2z\right)$ denotes the first derivative
of the polynomial $L_{n}\left(2z\right)$ with respect to the variable
$z$. The integral over the complete positive real axis shows that
the second entry condition of Theorem 2, Eq.~(\ref{eq:Entry Sum is One}),
is satisfied:

\begin{eqnarray}
\intop_{0}^{+\infty}\left(-1\right)^{n}\left(\delta\left(z\right)+\;L_{n}^{\prime}\left(2z\right)e^{-z}\right)\mathrm{d}z=1.\label{eq:Normalization Laguerre}
\end{eqnarray}

The nonnegativity of the integral in Eq. (\ref{eq:Entry Condition with Laguerre})
can be verified by using the recursion relation obtained as follows:\begin{subequations}

\begin{eqnarray}
\intop_{x}^{+\infty}\;L_{n}^{\prime}\left(2z\right)e^{-z}\mathrm{d}z & = & \intop_{x}^{+\infty}\;L_{n-1}^{\prime}\left(2z\right)e^{-z}\mathrm{d}z-2\intop_{x}^{+\infty}\;L_{n-1}\left(2z\right)e^{-z}\mathrm{d}z,\label{eq:Entry Step1}\\
 & = & -\intop_{x}^{+\infty}\;L_{n-1}^{\prime}\left(2z\right)e^{-z}\mathrm{d}z+2\intop_{x}^{+\infty}\;\left(L_{n-1}^{\prime}\left(2z\right)e^{-z}-L_{n-1}\left(2z\right)e^{-z}\right)\mathrm{d}z,\nonumber \\
 & = & -\intop_{x}^{+\infty}\;L_{n-1}^{\prime}\left(2z\right)e^{-z}\mathrm{d}z-2\,L_{n-1}\left(2x\right)e^{-x}.\label{eq:Entry Step3}
\end{eqnarray}
\end{subequations}where I have used the relation $L_{n}^{\prime}\left(x\right)=L_{n-1}^{\prime}\left(x\right)-L_{n-1}\left(x\right)$
applicable to a sequence of Laguerre polynomials in Eq.~(\ref{eq:Entry Step1})
and integrated by parts in Eq.~(\ref{eq:Entry Step3}). By making
use of this recursion, we eventually get

\begin{eqnarray}
\intop_{x}^{+\infty}\;\left(-1\right)^{n}\;L_{n}^{\prime}\left(2z\right)e^{-z}\mathrm{d}z & = & 2\,\sum_{i=0}^{n-1}\,\left(-1\right)^{i}\,L_{i}\left(2x\right)e^{-x}.\label{eq:Mixture of Fock states}
\end{eqnarray}
The integral in Eq. (\ref{eq:Entry Condition with Laguerre}) is proportional
to the radial profile of the Wigner function of an equally weighted
mixture of the first $n-1$ Fock states with the radial coordinate
$r=x^{\nicefrac{1}{2}}$. By Lemma 2, this quantity is nonnegative
$\forall x\geqslant0$. This means that each function $\tilde{W}_{n}(z)=\frac{\left(-1\right)^{n}}{\pi}\:L_{n}\left(2z\right)\;e^{-z}$
can be retrieved as the convolution of $e^{-z}$ with a generalized
function that meets the entry conditions of Theorem 2, Eqs.~(\ref{eq:Entry Partial Sum})
and (\ref{eq:Entry Sum is One}).

We note also - as I did already in Lemma 1 - that if various functions
fulfill Eqs.~(\ref{eq:Entry Partial Sum}) and (\ref{eq:Entry Sum is One}),
then a convex combination of those functions also fulfills the entry
conditions of Theorem 2 - including in the case where the convex combination
comprises an infinite number of such functions.

My final result is then a direct corollary of Eqs.~(\ref{eq:Fock State Definition}),
(\ref{eq:Link between W-majorization and majorization}) and Theorem
2: if the Wigner function of a mixture of Fock states is nonnegative,
then it is majorized by the Wigner function of the vacuum.

\section{Proofs of Theorem 1 and Theorem 2\label{sec:Proofs Theorems 1=0000262}}

\subsection*{Proof of Theorem 1 \label{sec:Proof of Theorem 1}}

Let $v{}_{0}=\left(a^{0},a^{1},a^{2},\dots,a^{N-1}\right){}^{\intercal}$
with $0<a<1$ and let $v{}_{k}$ and $u{}_{k},\;k=1,\dots,N$ be the
$N$-dimensional vectors defined by the elementary convolution

\medskip{}
$ $

\begin{eqnarray}
\left(\begin{array}{c}
v_{k}\\
u_{k}
\end{array}\right)=\left(\begin{array}{rrrrr}
0 & 0 & \cdots & 0 & 0\\
\vdots & \vdots &  & \vdots & \vdots\\
0 & \vdots &  &  & \vdots\\
\overset{\vphantom{10}}{-1} & 0\\
1 & -1 & \ddots & \vdots\\
0 & 1 & \ddots & 0 & \vdots\\
\vdots & 0 & \ddots & -1 & 0\\
 & \vdots & \ddots & 1 & -1\\
 &  &  & 0 & \overset{\vphantom{10}}{1}\\
\vdots &  &  & \vdots & 0\\
\vdots & \vdots &  & \vdots & \vdots\\
0 & 0 & \cdots & 0 & 0
\end{array}\right)\left(\begin{array}{c}
a^{0}\\
a^{1}\\
\vdots\\
\\
a^{N-1}
\end{array}\right)=\left(\begin{array}{c}
0\\
\vdots\\
0\\
\overset{\vphantom{10}}{-a^{0}}\\
\overset{\vphantom{10}}{a^{0}-a^{1}}\\
\vdots\\
\\
\vdots\\
a^{N-2}-a^{N-1}\\
\overset{}{\overset{}{a^{N-1}}}\\
\overset{\vphantom{10}}{0}\\
\vdots\\
0
\end{array}\right),\label{eq:Main Theorem-Vectors}
\end{eqnarray}
where $-a^{0}$ and $a^{N-1}$ are respectively the $k$-th and the
$k+N$-th elements of the vector {\footnotesize{}$\left(\begin{array}{c}
v_{k}\\
u_{k}
\end{array}\right)$}. The convolution $G$ is defined as

\begin{eqnarray}
G=\left(\begin{array}{c}
V\\
U
\end{array}\right)=\left(\begin{array}{c}
v_{0}\\
\bar{0}
\end{array}\right)+\sum_{k=1}^{N}\lambda_{k}\left(\begin{array}{c}
v_{k}\\
u_{k}
\end{array}\right), &  & \qquad\lambda_{k}\geqslant0,\label{eq:Main Theorem-Convolution}
\end{eqnarray}

where $\bar{0}$ denotes the vector of N zero values. Theorem 1 states
that

\begin{eqnarray}
\text{If }\;\left(\begin{array}{c}
V\\
U
\end{array}\right)\geqslant0\;\text{ Then }\;\left(\begin{array}{c}
v_{0}\\
\bar{0}
\end{array}\right)\succ\left(\begin{array}{c}
V\\
U
\end{array}\right).\label{eq:Main Theorem}
\end{eqnarray}

For the sake of clarity and to provide a geometric interpretation
of the theorem, I first present the proof for $N=3$.

Depending on the context, I use the words vector and point interchangeably
and similarly for the words elements and coordinates (of a point).

Because $\lambda_{k}\geqslant0$ in Eq.~(\ref{eq:Main Theorem-Convolution}),
the vector $U$ is always nonnegative and the nonnegativity of {\footnotesize{}${\scriptstyle \left(\begin{array}{c}
V\\
U
\end{array}\right)}$} is determined by the elements of the vector $V$.

\begin{figure}[H]
\noindent \begin{centering}
\includegraphics[scale=1.4]{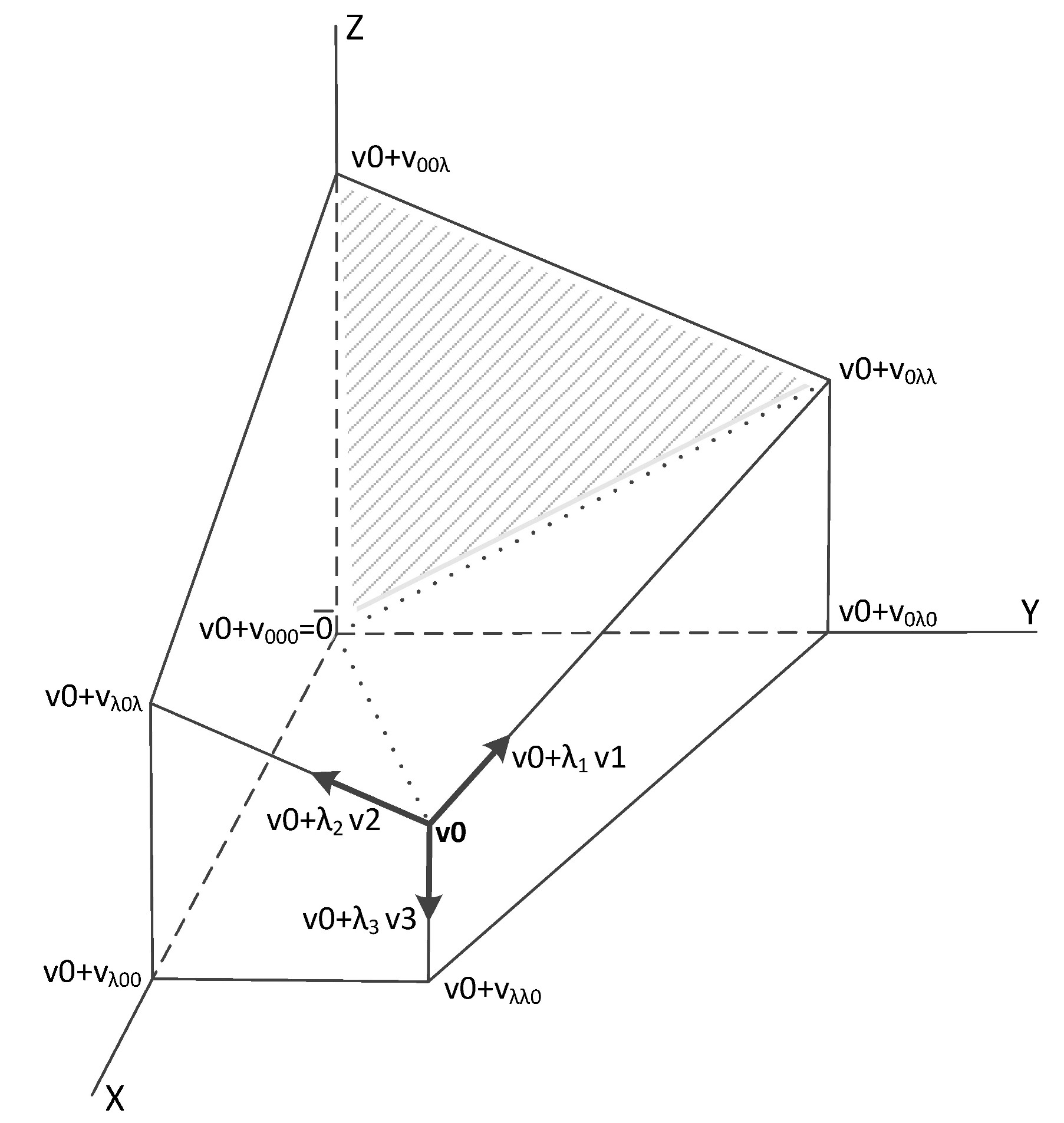}
\par\end{centering}

\caption{\label{fig:Polytope}$3$-dimensional convex cone spanned by the three
vectors $v{}_{1}$, $v{}_{2}$ and $v{}_{3}$ and cut by the planes
$X=0$, $Y=0$ and $Z=0$. Any point $V$ with positive coordinates
in this convex cone corresponds to a $6$-dimensional vector {\footnotesize{}${\scriptstyle \left(\protect\begin{array}{c}
V\protect\\
U
\protect\end{array}\right)}$} majorized by {\footnotesize{}$\left(\protect\begin{array}{c}
v_{0}\protect\\
\bar{0}
\protect\end{array}\right)$}.}
\end{figure}

In $\mathbb{R}{}^{3}$, the point $V$ is in a convex cone $C$ whose
vertex is in $v{}_{0}$ and that is defined by the three linearly
independent vectors $v_{1}$, $v_{2}$ and $v_{3}$. This convex cone
and its position with respect to the axes $X$, $Y$ and $Z$ are
schematized in Figure \ref{fig:Polytope}. 

A given point $V$ in $\mathbb{R}{}^{3}$ uniquely defines a point
{\footnotesize{}${\scriptstyle \left(\begin{array}{c}
V\\
U
\end{array}\right)}$} in $\mathbb{R}{}^{6}$ via Eqs.~(\ref{eq:Main Theorem-Vectors})
and (\ref{eq:Main Theorem-Convolution}). The subset of points $V$
of the convex cone $C$ such that {\footnotesize{}$\left(\begin{array}{c}
v_{0}\\
\bar{0}
\end{array}\right)\succ\left(\begin{array}{c}
V\\
U
\end{array}\right)$} is convex; this is a direct consequence of the convexity of majorization
\cite{Marshall&Olkin}.

The proof proceeds in two steps. In a first step, I decompose the
convex cone $C$ into six convex cones. In each of those six convex
cones, one of the plane $X=0$, $Y=0$ or $Z=0$ separates the points
with nonnegative coordinates from the points that have at least one
negative coordinate. In the second step, I complete the proof by showing
that $v{}_{0}$ majorizes each vertices of the six convex cones cut
by one of the plane $X=0$, $Y=0$ or $Z=0$.

Step1. To explain the first step, I replace the vectors $v_{1}$,
$v_{2}$ and $v_{3}$ by the proportional vectors $v_{0\lambda\lambda}$,
$v_{\lambda0\lambda}$ and $v_{\lambda\lambda0}$ respectively. The
example of the vector $v_{\lambda0\lambda}$ can be used to describe
how the indices are chosen: the $0$ as second index in $v_{\lambda0\lambda}$
indicates that the second coordinate of $v_{0}+v_{\lambda0\lambda}$
is zero whereas the $\lambda$ as 1\textsuperscript{st} and 3\textsuperscript{rd}
indices of $v_{\lambda0\lambda}$ indicate that for this vector, $\lambda_{1}=\lambda_{3}=0$
in Eq.~(\ref{eq:Main Theorem-Convolution}). This means that the
1\textsuperscript{st} and 3\textsuperscript{rd} coordinates of $v_{0}+v_{\lambda0\lambda}$
are nonnegative as a direct consequence of Eq.~(\ref{eq:Main Theorem-Vectors}).
The points $v_{0}+v_{iii}$ with $i=0$ or $i=\lambda$, are illustrated
in Figure \ref{fig:Polytope}.

Let $V$ be a given point of the convex cone C in $\mathbb{R}{}^{3}$,

\begin{equation}
V=v_{0}+\lambda_{1}^{\prime}v_{0\lambda\lambda}+\lambda_{2}^{\prime}v_{\lambda0\lambda}+\lambda_{3}^{\prime}v_{\lambda\lambda0},\qquad\lambda_{k}^{\prime}\geqslant0.\label{eq:Cone3 step1}
\end{equation}

As can be visualized in Figure \ref{fig:Polytope} and checked in
Table \ref{tab:Polytope}, the vector $\bar{0}=v_{0}+v_{000}$ is
within the convex cone $C$ and can be used to decompose it into three
convex cones. I assume without loss of generality that $V$ is in
the convex cone $C_{3}$ defined by the three vectors $v_{0\lambda\lambda}$,
$v_{\lambda0\lambda}$ and $v_{000}$,

\begin{equation}
V=v_{0}+\lambda_{1}^{\prime\prime}v_{0\lambda\lambda}+\lambda_{2}^{\prime\prime}v_{\lambda0\lambda}+\lambda_{3}^{\prime\prime}v_{000},\qquad\lambda_{k}^{\prime\prime}\geqslant0.\label{eq:Cone3 step2}
\end{equation}

I further decompose the convex cone $C_{3}$ by introducing the linear
combination of the first two vectors that span the cone, $v_{00\lambda}=\mu_{1}v_{0\lambda\lambda}+\mu_{2}v_{\lambda0\lambda}$,
where the two first coordinates of $v_{0}+v_{00\lambda}$ equal to
zero and, for this vector, $\lambda_{3}=0$ in Eq.~(\ref{eq:Main Theorem-Convolution}).
It will be clear in Table \ref{tab:Polytope} that the vector $v_{0}+v_{00\lambda}$
is within the convex cone $C_{3}$ (both $\mu_{1}$ and $\mu_{2}$
are nonnegative) and can be used to decompose $C_{3}$ into two convex
cones. I assume without loss of generality that the vector $v_{\lambda0\lambda}$
can be replaced in Eq.~(\ref{eq:Cone3 step2}) so that $V$ is in
the convex cone $C{}_{31}$ defined as 

\begin{equation}
V=v_{0}+\lambda_{1}^{\prime\prime\prime}v_{0\lambda\lambda}+\lambda_{2}^{\prime\prime\prime}v_{00\lambda}+\lambda_{3}^{\prime\prime}v_{000},\qquad\lambda_{k}^{\prime\prime\prime}\geqslant0.\label{eq:Cone3 step3}
\end{equation}

In summary, the initial convex cone $C$ can be decomposed into six
convex cones. The initial convex cone $C$ is the union of the six
convex cones and any point of the convex cone $C$ can be retrieved
in one of the six convex cones. Each of the six convex cones has properties
similar to $C{}_{31}$:
\begin{itemize}
\item The convex cone $C{}_{31}$ is defined by the three vectors $v_{0\lambda\lambda}$,
$v_{00\lambda}$ and $v_{000}$, denoted below by $v_{0ii}$ where
$i=0$ or $i=\lambda$;
\item by construction, each of the three vectors $v_{0}+v_{0ii}$ has the
first coordinate $X=0$ and a nonnegative value for the two other
coordinates;
\item as a consequence, within the convex cone $C{}_{31}$, the plane $X=0$
separates the points that have all nonnegative coordinates from the
points that have at least the first coordinate $X<0$. The hatched
area in Figure \ref{fig:Polytope} illustrates this separation for
the convex cone $C{}_{31}$.
\end{itemize}
Step 2. The second step of the proof consists in proving that the
vertices of the six convex cones limited by one of the plane $X=0$,
$Y=0$ or $Z=0$ correspond to vectors {\footnotesize{}${\scriptstyle \left(\begin{array}{c}
V\\
U
\end{array}\right)}$} that are all majorized by $v{}_{0}$. 

Table \ref{tab:Polytope} shows the coordinates of each of the points
$v_{0}+v_{iii}$ in $\mathbb{R}{}^{3}$, the corresponding values
$\lambda_{1}$, $\lambda_{2}$ and $\lambda_{3}$ as well as the corresponding
vectors {\footnotesize{}${\scriptstyle \left(\begin{array}{c}
V\\
U
\end{array}\right)}$} in $\mathbb{R}{}^{6}$.

\begin{table}[H]
\noindent \begin{centering}
\begin{tabular}{|>{\centering}m{1.9cm}|>{\centering}m{1.9cm}|>{\centering}m{4.5cm}|>{\centering}m{1cm}|>{\centering}m{1cm}|>{\centering}m{1cm}|}
\hline 
\noalign{\vskip\doublerulesep}
$v_{0}+v_{iii}$  & $V^{\intercal}$ & $\left(V^{\intercal},U^{\intercal}\right)$  & $\lambda{}_{1}$ & $\lambda{}_{2}$ & $\lambda{}_{3}$\tabularnewline[\doublerulesep]
\hline 
\hline 
\noalign{\vskip\doublerulesep}
$v{}_{0}$ & $\left(a^{0},a^{1},a^{2}\right)$ & $\left(a^{0},a^{1},a^{2},0,0,0\right)$ & $0$ & $0$ & $0$\tabularnewline[\doublerulesep]
\hline 
\noalign{\vskip\doublerulesep}
$v{}_{0}+v_{000}$ & $\left(0,0,0\right)$ & $\left(0,0,0,a^{0},a^{1},a^{2}\right)$ & $1$ & $1$ & $1$\tabularnewline[\doublerulesep]
\hline 
\noalign{\vskip\doublerulesep}
$v{}_{0}+v_{0\lambda\lambda}$ & $\left(0,a^{0},a^{1}\right)$ & $\left(0,a^{0},a^{1},a^{2},0,0\right)$ & $1$ & $0$ & $0$\tabularnewline[\doublerulesep]
\hline 
\noalign{\vskip\doublerulesep}
$v{}_{0}+v_{\lambda0\lambda}$ & $\left(a^{0},0,a^{1}\right)$ & $\left(a^{0},0,a^{1},a^{2}-a^{3},a^{3},0\right)$ & $0$ & $a{}^{1}$ & $0$\tabularnewline[\doublerulesep]
\hline 
\noalign{\vskip\doublerulesep}
$v{}_{0}+v_{\lambda\lambda0}$ & $\left(a^{0},a^{1},0\right)$ & $\left(a^{0},a^{1},0,a^{2}-a^{3},a^{3}-a^{4},a^{4}\right)$ & $0$ & $0$ & $a^{2}$\tabularnewline[\doublerulesep]
\hline 
\noalign{\vskip\doublerulesep}
$v{}_{0}+v_{\lambda00}$ & $\left(a^{0},0,0\right)$ & $\left(a^{0},0,0,a^{1}-a^{3},a^{2},a^{3}\right)$ & $0$ & $a^{1}$ & $a^{1}$\tabularnewline[\doublerulesep]
\hline 
\noalign{\vskip\doublerulesep}
$v{}_{0}+v_{0\lambda0}$ & $\left(0,a^{0},0\right)$ & $\left(0,a^{0},0,a^{1},a^{2}-a^{3},a^{3}\right)$ & $1$ & $0$ & $a^{1}$\tabularnewline[\doublerulesep]
\hline 
\noalign{\vskip\doublerulesep}
$v{}_{0}+v_{00\lambda}$ & $\left(0,0,a^{0}\right)$ & $\left(0,0,a^{0},a^{1},a^{2},0\right)$ & $1$ & $1$ & $0$\tabularnewline[\doublerulesep]
\hline 
\end{tabular}
\par\end{centering}

\caption{Values of $V^{\intercal}$ and corresponding $\left(V^{\intercal},U^{\intercal}\right)$
for the vertices of the polytope defined by the convex cone $C$ cut
by the planes $X=0$, $Y=0$ and $Z=0$.\label{tab:Polytope} }
\end{table}

It is directly apparent in Table \ref{tab:Polytope} that the eight
vectors {\footnotesize{}${\scriptstyle \left(\begin{array}{c}
V\\
U
\end{array}\right)}$} are majorized by the first vector {\footnotesize{}$\left(\begin{array}{c}
v_{0}\\
\bar{0}
\end{array}\right)$}. The first vector {\footnotesize{}$\left(\begin{array}{c}
v_{0}\\
\bar{0}
\end{array}\right)$} indeed majorizes itself and each other vector {\footnotesize{}${\scriptstyle \left(\begin{array}{c}
V\\
U
\end{array}\right)}$} may be obtained from {\footnotesize{}$\left(\begin{array}{c}
v_{0}\\
\bar{0}
\end{array}\right)$} by performing Robin Hood transfers.

The convexity property mentioned earlier completes the proof of the
theorem in the case $N=3$.

The following three observations remain valid for any value $N$.
\begin{itemize}
\item a majorized vector {\footnotesize{}${\scriptstyle \left(\begin{array}{c}
V\\
U
\end{array}\right)}$} can be the convolution of $v_{0}$ with a vector that has negative
elements. As an example, the point $v_{0}+v_{\lambda0\lambda}$ corresponds
to the convolution of $\left(a^{0},-a^{1},a^{1}\right)^{\intercal}$
with $v_{0}$;
\item a majorized vector {\footnotesize{}${\scriptstyle \left(\begin{array}{c}
V\\
U
\end{array}\right)}$} can correspond to the convex combination of vectors of the convex
cone $C$ that have negative elements. As an example, the point $v_{0}+v_{00\lambda}$
is the convex combination with equal weights of $v_{0}+2v_{1}$ and
$v_{0}+2v_{2}$ that both contain negative elements;
\item it is not necessary for the convolution to have the form specified
by Eq.~(\ref{eq:Main Theorem-Convolution}) to get a vector majorized
by $v{}_{0}$. As an example, the convolution with $\lambda_{1}=a$,
$\lambda_{2}=-\epsilon$ and $\lambda_{3}=a^{2}$ with $0<\epsilon\leqslant\left(1-a\right)^{2}$
defines a point outside of the convex cone $C$ that corresponds to
a vector {\footnotesize{}${\scriptstyle \left(\begin{array}{c}
V\\
U
\end{array}\right)}$} majorized by $v{}_{0}$.
\end{itemize}
The same two steps of the proof for $N=3$ directly generalize to
other values for $N$.

Step 1. The step by step decomposition of the convex cone generalizes
in the $N$-dimensional space into a decomposition that will end up
with $N!$ convex subcones.

A point of the initial convex cone $C$ is defined as\vspace{-10bp}

\begin{subequations}
\begin{align}
V & =v_{0}+\lambda_{1}v_{1}\hspace{16bp}+\lambda_{2}v_{2}\hspace{22bp}+\dots+\lambda_{N-1}v_{N-1}\hspace{7bp}+\lambda_{N}v_{N},\hspace{34bp}\lambda_{k}\geqslant0,\label{eq:ConeN step0a}\\
 & =v_{0}+\lambda_{1}^{\prime}v_{0\lambda\dots\lambda}+\lambda_{2}^{\prime}v_{\lambda0\lambda\dots\lambda}+\dots+\lambda_{N-1}^{\prime}v_{\lambda\dots\lambda0\lambda}+\lambda_{N}^{\prime}v_{\lambda\dots\lambda0},\qquad\lambda_{k}^{\prime}\geqslant0,\label{eq:ConeN step0b}
\end{align}
\end{subequations}where, in Eq.~(\ref{eq:ConeN step0b}), the $N$
indices uniquely define each vector $v_{i_{1}i_{2}\dots i_{N}}$ and
are chosen as in the case $N=3$: an index $i_{k}=0$ indicates that
the $k$-th coordinate of $v_{0}+v_{i_{1}i_{2}\dots i_{N}}$ is zero
and an index $i_{k}=\lambda$ indicates that $\lambda_{k}=0$ when
this vector is expressed in the initial convex cone defined by Eq.~(\ref{eq:Main Theorem-Convolution}).

It will be shown in Step 2 that all vectors $v_{0}+v_{i_{1}i_{2}\dots i_{N}}$,
with any combination of the indices, are inside the convex cone $C$.
This property, added to the fact that the $N$ vectors $v_{k}$ are
linearly independent, allows us to express any vector of the initial
convex cone as a vector of one of the $N!$ subcones that has, without
loss of generality, the following form:

\begin{equation}
V=v_{0}+\lambda_{1}^{\prime\prime}v_{0\lambda\dots\lambda}+\lambda_{2}^{\prime\prime}v_{00\lambda\dots\lambda}+\dots+\lambda_{N-1}^{\prime\prime}v_{0\dots0\lambda}+\lambda_{N}^{\prime\prime}v_{0\dots0},\qquad\lambda_{k}^{\prime\prime}\geqslant0,\label{eq:ConeN step1}
\end{equation}
where the first vector has one index $i_{1}=0$, the second vector
has two indices $i_{1}=i_{2}=0$, up to the last vector with $N$
zero indices $v_{0\dots0}$.

Similarly to the $N=3$ case, the following statements apply to the
$N!$ convex subcones (the indices and coordinates must be permuted
to describe properly each subcone):
\begin{itemize}
\item The $N$-dimensional convex subcone is defined by its vertex $v{}_{0}$
and by the $N$ vectors $v_{0\lambda\dots\lambda}$,..., $v_{0\dots0\lambda}$
and $v_{0\dots0}$;
\item the $N$ vectors $v_{0}+v_{0\lambda\dots\lambda}$,..., $v_{0}+v_{0\dots0\lambda}$
and $v_{0}+v_{0\dots0}$ have a first coordinate $X_{1}=0$ and a
nonnegative value for the other coordinates;
\item as a consequence and by linearity, within the convex subcone defined
by the $N$ vectors $v_{0\lambda\dots\lambda}$,..., $v_{0\dots0\lambda}$
and $v_{0\dots0}$, the hyperplane $X_{1}=0$ separate the points
that have all nonnegative coordinates from the points that have at
least the first coordinate $X_{1}<0$.
\end{itemize}
Step 2. Because the initial convex cone $C$ is the union of $N!$
convex subcones similar to the subcone defined by Eq.~(\ref{eq:ConeN step1}),
it must be proved that the vectors $v_{0}+v_{i_{1}i_{2}\dots i_{N}}$
with all possible combinations of the indices $i_{k}=0$ or $i_{k}=\lambda$
correspond to $2N$-dimensional vectors {\footnotesize{}${\scriptstyle \left(\begin{array}{c}
V\\
U
\end{array}\right)}$} that  are majorized by $v{}_{0}$.

I use a step by step procedure to construct the points $v_{0}+v_{i_{1}i_{2}\dots i_{N}}$
of the convex cone $C$ such that either $i_{k}=0$ or $i_{k}=\lambda$,
starting from the lowest index.

Let us assume that the $j$-th element is the first (lowest index)
zero of a given vector $v_{0}+v_{i_{1}i_{2}\dots i_{N}}$: $i{}_{j}=0$
and $i_{1}=\dots=i_{j-1}=\lambda$. We can observe in Eq.~(\ref{eq:Main Theorem-Vectors})
that any vector $v_{k}$ has only one negative element, its $k$-th
element, whose value is $-a^{0}=-1$.

The initial value of the $j$-th element of $v{}_{0}$ is $a^{j-1}$
and - within the convex cone $C$ - a zero can only be obtained by
adding a vector $a^{j-1}v_{j}$ to $v_{0}$. Using Eq.~(\ref{eq:Main Theorem-Vectors}),
the effects of adding $a^{j-1}${\footnotesize{}$\left(\begin{array}{c}
v_{j}\\
u_{j}
\end{array}\right)$} to {\footnotesize{}$\left(\begin{array}{c}
v_{0}\\
\bar{0}
\end{array}\right)$} can be described as follows:
\begin{itemize}
\item The first $j-1$ elements of $V$ are unchanged and keep their values
$a^{0},a^{1},\dots,a^{j-2}$;
\item the $j$-th element of $V$ decreases from the value $a^{j-1}$ to
$0$; we can consider that this element decreases by slices, that
is, first from $a^{j-1}$ to $a^{j}$, then from $a^{j}$ to $a^{j+1}$,
and so on, until the $N$-th slice, from $a^{N+j-2}$ to $0$;
\item the decrease in value of the $j$-th element of $V$ is compensated
by increases of the values of the $N$-th subsequent elements of{\footnotesize{}
${\scriptstyle \left(\begin{array}{c}
V\\
U
\end{array}\right)}$}: the $j+1$-th element is increased by the first slice, that is,
from $a^{j}$ to $a^{j-1}$, the next element is increased by the
second slice from $a^{j+1}$ to $a^{j}$ and so on. Those transfers
within the $V$ part of {\footnotesize{}${\scriptstyle \left(\begin{array}{c}
V\\
U
\end{array}\right)}$} are all Robin Hood transfers;
\item after the transfers to the $V$ part of {\footnotesize{}${\scriptstyle \left(\begin{array}{c}
V\\
U
\end{array}\right)}$} are completed, the subsequent slices are transferred to the $U$
part of{\footnotesize{} ${\scriptstyle \left(\begin{array}{c}
V\\
U
\end{array}\right)}$}. Before these transfers, all the elements of $U$ had a zero value,
which means that all the transfers from the $j$-th element of $V$
to the $U$ part of {\footnotesize{}${\scriptstyle \left(\begin{array}{c}
V\\
U
\end{array}\right)}$} are also Robin Hood transfers;
\item after this first step, the $N$ elements of {\footnotesize{}${\scriptstyle \left(\begin{array}{c}
V\\
U
\end{array}\right)}$} that have been increased ($j+1$-th to $j+N$-th) have the form
$a^{i}$ or $a^{i}-a^{i+p}$ where $p$ is a positive integer and
where the exponent $i$ increases continuously from element to element,
by steps equal to $1$.
\end{itemize}
The effect of adding an additional vector $a^{j^{\prime}-2}${\footnotesize{}$\left(\begin{array}{c}
v_{j^{\prime}}\\
u_{j^{\prime}}
\end{array}\right)$} to the obtained vector can be sliced into Robin Hood transfers in
the same way as for the first vector. After each addition, the vector{\footnotesize{}
${\scriptstyle \left(\begin{array}{c}
V\\
U
\end{array}\right)}$} keeps the same structure and in particular, the elements that follow
the element that has just been put to zero keep the same structure
$a^{i}$ or $a^{i}-a^{i+p}$ where $p$ is a positive integer and
where the exponent $i$ increases from element to element, by steps
equal to $1$.

Two steps of this procedure are illustrated in Table \ref{tab:Steps for convolutions}
to construct the vector $v_{0}+v_{\lambda\lambda0\lambda\lambda0\lambda}$
in the case $N=7$, first for $j=3$ and then for $j^{\prime}=6$.

\begin{table}[H]
\noindent \begin{centering}
\[
\begin{array}{ccccc}
\begin{array}{c}
v_{0}+v_{\lambda\lambda\lambda\lambda\lambda\lambda\lambda}\end{array} & \longrightarrow & v_{0}+v_{\lambda\lambda0\lambda\lambda\lambda\lambda} & \longrightarrow & v_{0}+v_{\lambda\lambda0\lambda\lambda0\lambda}\\
\\
\\
\left(\begin{array}{c}
v_{0}\\
\bar{0}
\end{array}\right) & \longrightarrow & \left(\begin{array}{c}
v_{0}\\
\bar{0}
\end{array}\right)+a^{2}\left(\begin{array}{c}
v_{3}\\
u_{3}
\end{array}\right) & \longrightarrow & \left(\begin{array}{c}
v_{0}\\
\bar{0}
\end{array}\right)+a^{2}\left(\begin{array}{c}
v_{3}\\
u_{3}
\end{array}\right)+a^{4}\left(\begin{array}{c}
v_{6}\\
u_{6}
\end{array}\right)\\
\\
\\
\left(\begin{array}{c}
\\
a{}^{0}\\
a{}^{1}\\
a{}^{2}\\
a{}^{3}\\
a{}^{4}\\
a{}^{5}\\
a{}^{6}\\
0\\
0\\
0\\
0\\
0\\
0\\
0\\
\\
\end{array}\right) & \longrightarrow & \left(\begin{array}{ccc}
\\
 & a{}^{0}\\
 & a{}^{1}\\
 & 0\\
 & a{}^{2}\\
 & a{}^{3}\\
 & a{}^{4}\\
 & a{}^{5}\\
 & a^{6}-a^{7}\\
 & a^{7}-a^{8}\\
 & a^{8}\\
 & 0\\
 & 0\\
 & 0\\
 & 0\\
\\
\end{array}\right) & \longrightarrow & \left(\begin{array}{ccc}
\\
 & a{}^{0}\\
 & a{}^{1}\\
 & 0\\
 & a{}^{2}\\
 & a{}^{3}\\
 & 0\\
 & a{}^{4}\\
 & a^{5}-a^{7}\\
 & a^{6}-a^{8}\\
 & a^{7}\\
 & a^{8}-a^{9}\\
 & a^{9}-a^{10}\\
 & a^{10}\\
 & 0\\
\\
\end{array}\right)
\end{array}
\]

\par\end{centering}

\caption{Illustration of the two successive steps to construct the vector $v_{0}+v_{\lambda\lambda0\lambda\lambda0\lambda}$
in the case $N=7$. It is clear that $v{}_{0}$ majorizes the vector{\footnotesize{}
${\scriptstyle \left(\protect\begin{array}{c}
V\protect\\
U
\protect\end{array}\right)}$} corresponding to $v_{0}+v_{\lambda\lambda0\lambda\lambda\lambda\lambda}$,
which in turn majorizes the vector {\footnotesize{}${\scriptstyle \left(\protect\begin{array}{c}
V\protect\\
U
\protect\end{array}\right)}$} corresponding to $v_{0}+v_{\lambda\lambda0\lambda\lambda0\lambda}$.
\label{tab:Steps for convolutions} }
\end{table}
The construction described above demonstrates how a vector $v_{0}+v_{i_{1}i_{2}\dots i_{N}}$
with any combination of the indices $i_{k}=0$ or $i_{k}=\lambda$,
can be obtained from the initial vector {\footnotesize{}$\left(\begin{array}{c}
v_{0}\\
\bar{0}
\end{array}\right)$} by successively adding vectors $\lambda_{k}${\footnotesize{}$\left(\begin{array}{c}
v_{k}\\
u_{k}
\end{array}\right)$} for each position $k$ where the index $i_{k}=0$ in $v_{0}+v_{i_{1}i_{2}\dots i_{N}}$.
Each addition of vector $\lambda_{k}${\footnotesize{}$\left(\begin{array}{c}
v_{k}\\
u_{k}
\end{array}\right)$} - with $\lambda{}_{k}$ properly chosen - can be decomposed into
Robin Hood transfers. Consequently, the vector {\footnotesize{}$\left(\begin{array}{c}
v_{0}\\
\bar{0}
\end{array}\right)$} majorizes all vectors{\footnotesize{} ${\scriptstyle \left(\begin{array}{c}
V\\
U
\end{array}\right)}$} constructed in this manner. This completes the proof of Theorem
1.

\subsection*{Proof of Theorem 2 \label{sec:Proof of Theorem 2}}

I outline (only) the main steps of the extension of Theorem 1 to the
continuous case. My approach consists in showing that the continuous
convolution of Theorem 2 can be obtained as the limit of a sequence
of vectors that match the entry criteria of Theorem 1.

Let $c(z)$ be the function defined in the statement of Theorem 2
in section \ref{sec:Theorem 2}.

I define an interval $\left[z_{0},z_{end}\right]$ on the positive
real axis with $z_{0}=0$ and I partition that interval into $2N$
equal sub-intervals such that $z{}_{i}-z_{i-1}=\Delta z,\:i=1,\dots,2N$.
Within the first $N$ sub-intervals $\left[z_{i-1},z{}_{i}\right],\:i=1,\dots,N,$
I choose $z_{i}^{\prime}$ so that $c\left(z_{i}^{\prime}\right)$
is the supremum of the function $c\left(z\right)$ on that sub-interval.

I construct a $N+1$-dimensional vector $C_{N+1}$ that fulfills (for
$\Delta z$ sufficiently small) the condition specified by Eq.~(\ref{eq:Sum of Last N}):

\begin{eqnarray}
C{}_{N+1}=\tilde{k}_{N+1}\left(\begin{array}{c}
c_{\delta0}\\
\overset{\vphantom{10}}{c\left(z_{1}^{\prime}\right)\left(1-e^{-\Delta z}\right)}\\
\vdots\\
c\left(z_{N-1}^{\prime}\right)\left(1-e^{-\Delta z}\right)\\
\overset{\vphantom{10}}{c_{last}\frac{\left(1-e^{-\Delta z}\right)}{\Delta z}}
\end{array}\right),\label{eq:Vector CN+1}
\end{eqnarray}

where $c_{\delta0}$ is the factor multiplying the Dirac delta function
of $c(z)$ in $z=0$ , $\tilde{k}_{N+1}$ ensures that the vector
$C{}_{N+1}$ is normalized and

\begin{equation}
c_{last}=\intop_{z_{N}}^{+\infty}\;c\left(z\right)\;\mathrm{d}z.\label{eq:C Last}
\end{equation}

I also define the $N$-dimensional vector $e{}_{0}=\left(e^{-0\Delta z},e^{-1\Delta z},e^{-2\Delta z},\dots,e^{-\left(N-1\right)\Delta z}\right)^{\intercal}$.

It can be seen that the discrete convolution $G_{2N}=C_{N+1}*e_{0}$
is nonnegative by decomposing the continuous convolution $g\left(z\right)=c\left(z\right)\ast e^{-z}$
into a discrete sum. It is calculated at one of the points that define
the first $N-1$ sub-intervals, $z=z_{j}$ as follows\vspace{-10bp}
\begin{subequations}

\begin{align}
g\left(z_{j}\right) & =\sum_{i=1}^{j}\;\intop_{z_{i-1}}^{z_{i}}c\left(x\right)e^{-\left(z_{j}-x\right)}\mathrm{d}x\label{eq:Continuous=000026Discrete1}\\
 & =c_{\delta0}\:e^{-z_{j}}+\sum_{i=1}^{j}\;c\left(z_{i}^{\prime\prime}\right)\;\intop_{z_{i-1}}^{z_{i}}e^{-\left(z_{j}-x\right)}\mathrm{d}x\label{eq:Continuous=000026Discrete2}\\
 & =c_{\delta0}\:e^{-z_{j}}+\left(1-e^{-\Delta z}\right)\;\sum_{i=1}^{j}\;c\left(z_{i}^{\prime\prime}\right)\;e^{-\left(z_{j}-z_{i}\right)},\label{eq:Continuous=000026Discrete3}
\end{align}

\end{subequations}where the points $z_{i}^{\prime\prime}$ within
each sub-intervals are determined by the mean value theorem.

The points $z_{i}^{\prime}$ were chosen so that $c\left(z_{i}^{\prime}\right)\geqslant c\left(z_{i}^{\prime\prime}\right),\:i=1,\dots,N$.
The sum in Eq.~(\ref{eq:Continuous=000026Discrete3}) has the same
form as the discrete convolution $G_{2N}$ and it shows that if the
continuous convolution $g\left(z_{j}\right)$ is nonnegative, then
the $j+1$-th element of the discrete convolution $G_{2N}$ is nonnegative
- also at the points where the continuous convolution $g\left(z\right)$
might have a zero value.

A second variable name $y$ is associated with the index $j$ and
the $j+1$-th element of the discrete convolution $G_{2N}=C_{N+1}*e_{0}$
is denoted $G_{2N}\left(y_{j}\right),\:j=0,\dots,2N-1$.

The three vectors $C_{N+1}$, $e_{0}$ and $G_{2N}$ are within the
conditions of Theorem 1 for any sufficiently small value of $\Delta z$
as well as for any choice of $z_{end}$ that defines the initial interval.

Because in this case $e_{0}\;\succ G_{2N}$, Eq.~(\ref{eq:Definition Discrete-Doubly Stochastic})
can be used to write the $2N$ equations

\begin{equation}
G_{2N}\left(y_{j}\right)=\sum_{i=0}^{N-1}D_{2N}\left(z_{i,},y_{j}\right)\;e^{-z_{i}},\qquad j=0,\dots,2N-1,\label{eq:Riemann-Stieltjes Sum}
\end{equation}
where $0\leqslant D_{2N}\left(z_{i,},y_{j}\right)\leqslant1$ are
the elements of a $2N\times2N$ doubly stochastic matrix.

As $\Delta z$ tends to zero, $N$ to infinity and $z_{end}$ to infinity,
\begin{itemize}
\item the vector $G_{2N}$ tends to the continuous convolution $g\left(z\right)$
of the distribution $e{}^{-z}$ with the function $c\left(z\right)$
(both are integrable functions and improper integrals are absolutely
convergent);
\item the sum in Eq.~(\ref{eq:Riemann-Stieltjes Sum}) takes the form of
a convergent Riemann\textendash Stieltjes integral whose integrand
is $e{}^{-z}$ and whose integrator function is monotone and of bounded
variation (this integrator function could be a generalized function).
\end{itemize}
At the limit, the sum in Eq.~(\ref{eq:Riemann-Stieltjes Sum}) identifies
with the condition for continuous majorization set out in Eq.~(\ref{eq:Definition Continuous Doubly Stochastic}).
This completes the proof of Theorem 2.

\section{Conclusion\label{sec:Conclusion}}

I have proved that the nonnegative Wigner function of any mixture
of Fock states is majorized by the Wigner function of the vacuum.

The fact that the Wigner function results from a convolution is revealed
to be a decisive factor in this majorization relation. I have described
the underlying new majorization property dedicated to convolutions
with the negative exponential distribution.

As a corollary to the majorization relation, the Shannon differential
entropy of the Wigner function of the vacuum is lower than the entropy
of the Wigner function of any mixture of Fock states. This same order
between the Wigner functions of the vacuum and any mixture of Fock
states is also respected for any other entropy, provided that the
latter is defined by a concave function.

I conjecture that properties similar to the ones described in this
article may explain further majorization relations between the Wigner
function of the vacuum and other states than mixtures of Fock states
- after all, any Wigner function can be obtained as a convolution
of the Sudarshan-Glauber P representation.

\bibliographystyle{unsrt}

\end{document}